\documentclass[journal]{IEEEtran}

\usepackage{graphicx}
\usepackage{epsfig}
\usepackage{dcolumn}
\usepackage{bm}
\usepackage{amsmath}
\usepackage{amssymb}

\begin{document}
%
\title{Memristive circuits simulate memcapacitors and meminductors}
%
%
%

\author{Yuriy~V.~Pershin and Massimiliano~Di~Ventra
\thanks{Yu. V. Pershin is with the Department of Physics
and Astronomy and USC Nanocenter, University of South Carolina,
Columbia, SC, 29208 \newline e-mail: pershin@physics.sc.edu.}
\thanks{M. Di Ventra is with the Department
of Physics, University of California, San Diego, La Jolla,
California 92093-0319 \newline e-mail: diventra@physics.ucsd.edu.}

}

%
%


\maketitle

\begin{abstract}
We suggest electronic circuits with
memristors (resistors with memory) that operate as memcapacitors
(capacitors with memory) and meminductors (inductors with memory).
Using a memristor emulator, the suggested circuits
have been built and their operation has been demonstrated, showing a useful and interesting
connection between the three memory elements.
\end{abstract}


%
\IEEEpeerreviewmaketitle

{\it Introduction:} Memcapacitive and meminductive systems are two
recently postulated classes of circuit elements with memory
\cite{DiVentra2009-2} that complement the class of memristive
systems \cite{Chua1976-1,Chua1971-1}. Their main characteristic is
a hysteretic loop - which may or may not pass through the
origin~\cite{DiVentra2009-2} - in their constitutive variables
(charge-voltage for memcapacitors and current-flux for
meminductors) when driven by a periodic input, and, unlike
memristors, they can store energy. As of today, a few systems have
been found to operate as memcapacitors and meminductors (see
\cite{DiVentra2009-2} and references therein). However, these are
neither available on the market yet, nor their properties can be
easily tuned to investigate their role in more complex circuits.
The same can be said about memristive systems. Therefore,
electronic emulators of such memory elements that could be easily
built and tuned would be highly desirable. We have previously
designed and built a {\em memristor emulator} and shown its use in
neuromorphic and programmable analog
circuits~\cite{Pershin2009-3,Pershin2009-4}. In this Letter, we
use such memristor emulator to design and build {\em memcapacitor}
and {\it meminductor emulators}, and prove experimentally their
main properties. Since all of these emulators can be built from
inexpensive off-the-shelf components we expect them to be
extremely useful in the design, understanding and simulations of
complex circuits with memory.
\begin{figure}[h]
 \begin{center}
\includegraphics[angle=0,width=6.0cm]{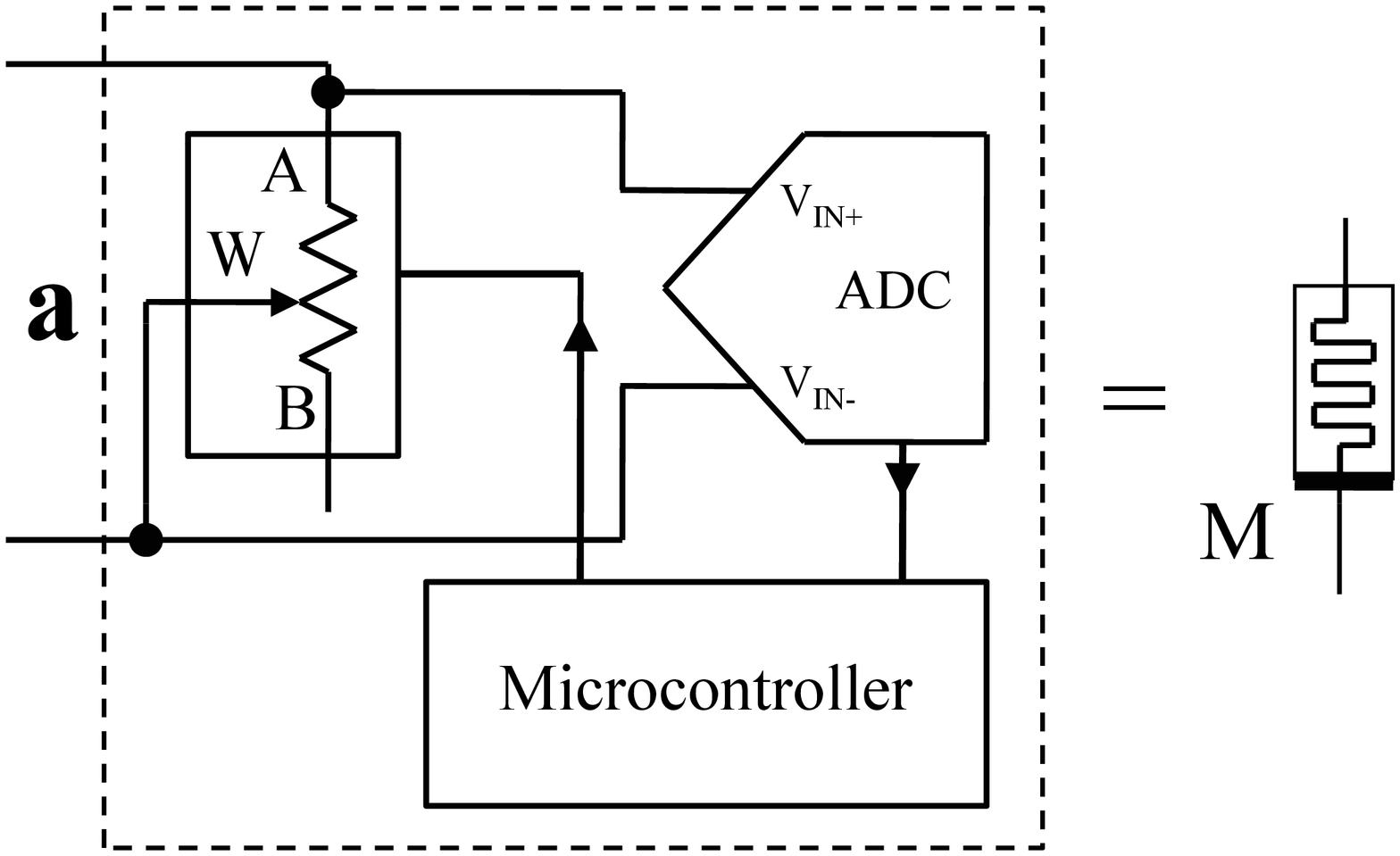}
\includegraphics[angle=-90,width=6.0cm]{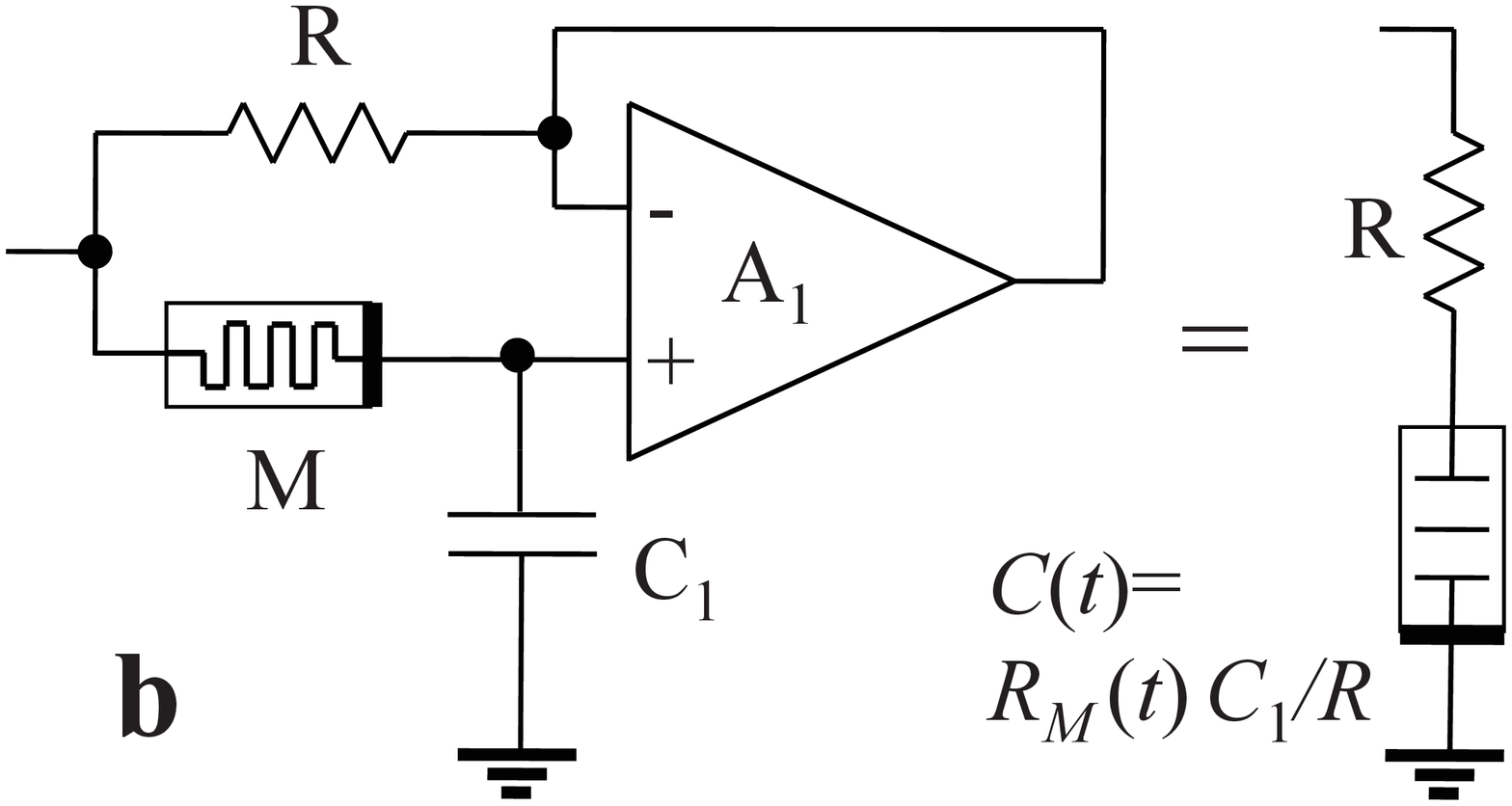}
\includegraphics[angle=-90,width=6.0cm]{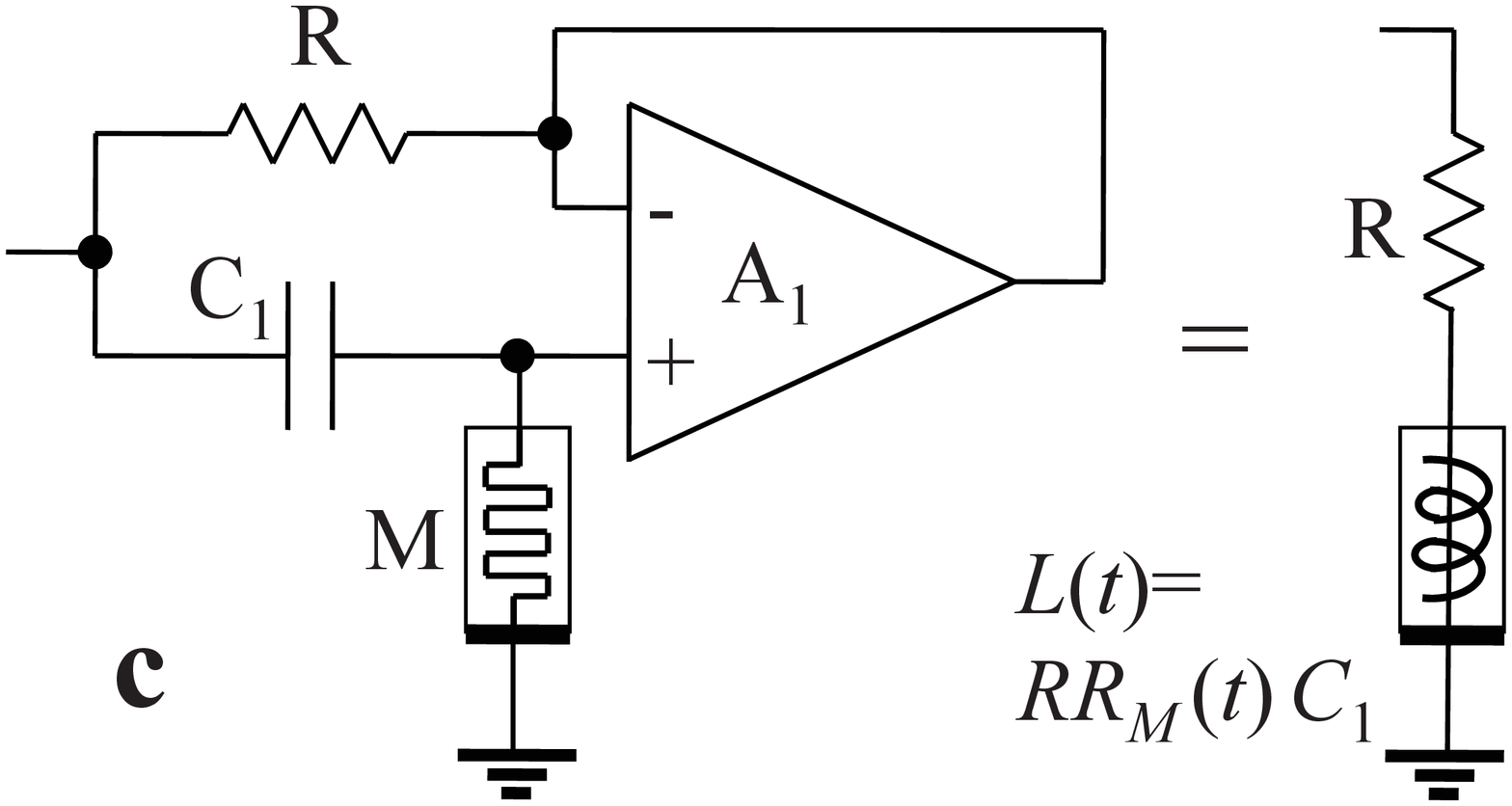}
\caption{\label{fig1} Circuits simulating (a) memristor, (b) memcapacitor and
(c) meminductor. Their approximate equivalent circuits are shown on the right.}
 \end{center}
\end{figure}

{\it Proposed circuits:} The proposed circuits of memcapacitor and
meminductor emulators are shown in Fig. \ref{fig1}(b) and (c),
respectively, together with the memristor emulator in Fig.
\ref{fig1}(a). The memristor emulator is implemented as in
Refs~\cite{Pershin2009-3,Pershin2009-4} and consists of a digital
potentiometer whose resistance is continuously updated by a
microcontroller and determined by pre-programmed equations of
current-controlled or voltage-controlled memristive systems. In
our experiments, the memristance $R_M$ is governed by an
activation-type model \cite{Pershin2009-4,Pershin2009-2} defined
by $R_M=x$ with
\begin{eqnarray}
\dot x&=&\left(\beta V_M+0.5\left( \alpha-\beta\right)\left[
|V_M+V_T|-|V_M-V_T| \right]\right) \nonumber\\ & &\times \theta
\left( x-R_{min}\right)\theta\left( R_{max}-x\right)
\label{Mmodel2},
\end{eqnarray}
where $V_M$ is the voltage across the memristor, and
$\theta(\cdot)$ is the step function. We choose for this work the
following parameters $\alpha=0$, $\beta=62$k$\Omega/$(V$\cdot$s)
(used in the memcapacitor emulator), $\beta=1$M$\Omega/$(V$\cdot$s)
(used in the meminductor emulator), $V_T=1$V, $R_{min}=5$k$\Omega$ and
$R_{max}=10$k$\Omega$.

The memcapacitor emulator consists of this memristor $M$, a
capacitor $C_1$ and a resistor $R$ connected to an operational
amplifier $A_1$ as shown in Fig. \ref{fig1}b. Since the
operational amplifier keeps nearly equal voltages at its positive
and negative inputs, the voltage on the capacitor $C_1$ is applied
to the right terminal of $R$. Therefore, we can think that an
effective capacitor with a time-dependent capacitance $C(t)$ is
connected to the right terminal of $R$, so that the relation
$RC(t)=R_M(t)C_1$ holds. (Note that the voltage at the capacitor
$V_C$ is equivalent to the voltage, $V_{-}$, at the negative
terminal of the operational amplifier.) This allows us to
determine the capacitance as $C(t)=R_M(t)C_1/R=(V_{in}-V_{-})/(R
dV_{-}/dt)$ since $R_M(t)=(V_{in}-V_{-})/I=(V_{in}-V_{-})/(C_1
dV_{-}/dt)$. In the limit $R \ll R_M$, we obtain the approximate
equivalent circuit shown on the right of Fig. \ref{fig1}b. On the
other hand, the meminductor emulator is similar to the design of a
gyrator with a memristor replacing a resistor, and the equivalent
inductance $L(t)=RR_M(t)C_1$, as it is evident from Fig.
\ref{fig1}c. In both cases, the time dependence of the equivalent
capacitance, $C$, and inductance, $L$, is due to the time
dependence of $R_M$.

\begin{figure}[h]
 \begin{center}
\includegraphics[angle=-90,width=7.5cm]{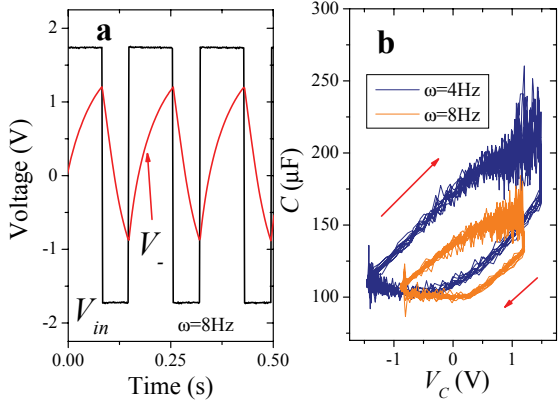}
\caption{\label{fig2} Memcapacitor emulator response to a square
wave signal. We used the circuit shown in Fig. \ref{fig1}b with
$R=480\Omega$ and $C_1=10\mu$F. {\bf a} Time-dependence of the
input voltage signal $V_{in}$ and voltage at the negative input of
the operational amplifier $\textnormal{A}_1$. {\bf b}. The
equivalent capacitance $C(t)$ numerically extracted from $V_{in}$
and $V_-$ signals as described in the text.}
 \end{center}
\end{figure}

\begin{figure}[h]
 \begin{center}
\includegraphics[angle=-90,width=7.5cm]{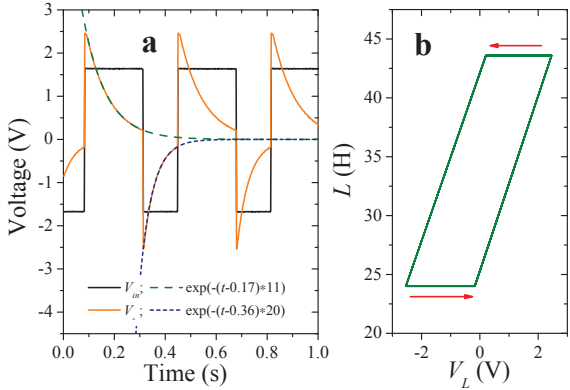}
\caption{\label{fig3} Meminductor emulator (Fig. \ref{fig1}c with
$R=480\Omega$ and $C_1=10\mu$F) response to a square wave signal.
{\bf a} Time-dependence of the input voltage signal $V_{in}$ and
voltage $V_{-}=V_L$ at the negative input of the operational amplifier
$\textnormal{A}_1$. {\bf b} Schematics of meminductor hysteresis
loop drawn with the inductance $L$ obtained using exponential fits to $V_-$
signals as shown in {\bf a}.}
 \end{center}
\end{figure}

In order to prove that these circuits emulate the behavior of
memcapacitors and meminductors, we have analyzed their response
under the application of a square wave signal. This is shown in
Fig. \ref{fig2}, where, on the left panel, we show both the input
voltage $V_{in}$ and the voltage at the negative terminal of the
operational amplifier $V_{-}$, and, on the right panel, the
equivalent capacitance of the memcapacitor emulator at two values
of frequency of the square wave signal. Clear hysteresis loops are
visible in the capacitance as a function of the voltage at the
capacitor $V_C=V_{-}$. We also note that the capacitance
hysteresis is frequency dependent: the loop is much smaller at the
higher frequency of 8Hz. This is a manifestation of a typical
property of memory circuit elements \cite{DiVentra2009-2} that at
high frequencies behave as linear elements. The fluctuations of
$C$ in Fig.~\ref{fig2}b are related to the limited resolution of
our data acquisition system and some noise in the circuit.

Similar considerations apply to the meminductor emulator as
demonstrated in Fig.~\ref{fig3}. Here, it is clearly seen that the
shape of the $V_-$ signal (which in this case is equal to the
voltage on the equivalent inductor $V_L$) depends on the polarity
of applied voltage. We extracted numerically the equivalent
inductance $L$ from $V_{in}$ and $V_-$ signals and found that it
contains a considerable amount of noise. Less noisy estimation of
equivalent inductance is obtained using a fit of $V_-$ signal by
exponentially decaying curves as demonstrated in Fig. \ref{fig3}a
giving a decaying time $\tau=L/R$, from which we have extracted
$L$. Since the memristor emulator state in the meminductor
emulator changes fast (its parameters are given below Eq.
(\ref{Mmodel2})), the equivalent inductance $L$ switches between
two limiting values as shown schematically in Fig. \ref{fig3}b.

Finally, we would like to mention that although the suggested
emulators reproduce the essential features of real memcapacitors and
meminductors, certain aspects are different. In particular, the
designed emulators are active devices requiring a power source for
their operation. More importantly, these emulators do not actually
store energy, which might be a limitation in specific
applications. However, from the point of view of circuit response,
almost any kind of memcapacitor and meminductor operation model
can be realized using an appropriate memristor-emulator operation
algorithm.

{\it Conclusions:} We have demonstrated that simple circuits with
memristors can exhibit both memcapacitive and meminductive
behavior. Memcapacitor and meminductor emulators have been
designed and built using the previously suggested memristor
emulator~\cite{Pershin2009-3,Pershin2009-4} since solid-state
memristors are not available yet. These emulators can be created
from inexpensive off-the-shelf components, and as such they
provide powerful tools to understand the different functionalities
of these newly suggested memory elements without the need of
expensive material fabrication facilities. We thus expect they
will be of use in diverse areas ranging from non-volatile memory
applications to neuromorphic circuits.

{\it Acknowledgment:} This work has been partially funded by the
NSF grant No. DMR-0802830.

\bibliographystyle{IEEEtran}
\bibliography{IEEEabrv,memristor}

\begin{thebibliography}{1}
\providecommand{\url}[1]{#1}
\csname url@samestyle\endcsname
\providecommand{\newblock}{\relax}
\providecommand{\bibinfo}[2]{#2}
\providecommand{\BIBentrySTDinterwordspacing}{\spaceskip=0pt\relax}
\providecommand{\BIBentryALTinterwordstretchfactor}{4}
\providecommand{\BIBentryALTinterwordspacing}{\spaceskip=\fontdimen2\font plus
\BIBentryALTinterwordstretchfactor\fontdimen3\font minus
  \fontdimen4\font\relax}
\providecommand{\BIBforeignlanguage}[2]{{%
\expandafter\ifx\csname l@#1\endcsname\relax
\typeout{** WARNING: IEEEtran.bst: No hyphenation pattern has been}%
\typeout{** loaded for the language `#1'. Using the pattern for}%
\typeout{** the default language instead.}%
\else
\language=\csname l@#1\endcsname
\fi
#2}}
\providecommand{\BIBdecl}{\relax}
\BIBdecl

\bibitem{DiVentra2009-2}
M.~Di~Ventra, Y.~V. Pershin, and L.~O. Chua, ``Circuit elements with memory:
  memristors, memcapacitors and meminductors,'' \emph{Proc. {IEEE}}, vol.~97,
  pp. 1717--1724, 2009.

\bibitem{Chua1976-1}
L.~O. Chua and S.~M. Kang, ``\BIBforeignlanguage{English}{Memristive devices
  and systems},'' \emph{\BIBforeignlanguage{English}{Proc. {IEEE}}}, vol.~64,
  no.~2, pp. 209--223, 1976.

\bibitem{Chua1971-1}
L.~O. Chua, ``\BIBforeignlanguage{English}{Memristor - the missing circuit
  element},'' \emph{\BIBforeignlanguage{English}{{IEEE} Trans. Circuit
  Theory}}, vol.~18, no.~5, pp. 507--519, 1971.

\bibitem{Pershin2009-3}
Y.~V. Pershin and M.~Di~Ventra, ``Experimental demonstration of associative
  memory with memristive neural networks,'' \emph{arXiv:0905.2935}, 2009.

\bibitem{Pershin2009-4}
------, ``Practical approach to programmable analog circuits with memristors,''
  \emph{arXiv:0908.3162}, 2009.

\bibitem{Pershin2009-2}
Y.~V. Pershin, S.~La~Fontaine, and M.~Di~Ventra, ``Memristive model of amoeba's
  learning,'' \emph{Phys. Rev. E}, vol.~80, p. 021926, 2009.

\end{thebibliography}
\end{document}